# LiNbO$_3$ ridge waveguides realized by precision dicing on silicon for high efficiency second harmonic generation

Mathieu Chauvet[1*], Fabien Henrot[1], Florent Bassignot[2], Fabrice Devaux[1], Ludovic Gauthier-Manuel[1], Vincent Pêcheur[1], Hervé Maillotte[1], Brahim Dahmani[3]

[1]*FEMTO-ST institute, UMR CNRS 6174, Université de Franche-Comté, 15B avenue des Montboucons, 25000 Besançon – France*
[2]*Femto-Engineering, 15B avenue des Montboucons, 25000 Besançon – France*
[3]*LOVALITE, 7 rue Xavier Marmier, 25000 Besançon - France*
[*]*mathieu.chauvet@univ-fcomte.fr*

**Abstract:** Nonlinear periodically poled ridge LiNbO$_3$ waveguides have been fabricated on silicon substrates. Components are carved with only use of a precision dicing machine without need for grinding or polishing steps. They show efficient second harmonic generation at telecommunication wavelengths with normalized conversion reaching 204%/W in a 15 mm long device. Influence of geometrical non uniformities of waveguides due to fabrication process is asserted. Components characteristics are studied notably their robustness and tunability versus temperature.

**References and links**

1. Yoo S. J. B., "Wavelength conversion technologies for WDM network applications", J. of Lightwave Technology, **14**, 955-966 (1996).
2. V. G. Ta'eed, M. R. E. Lamont, D. J. Moss, B. J. Eggleton, D. Choi, S. Madden, B. Luther-Davies, "All optical wavelength conversion via cross phase modulation in chalcogenide glass rib waveguides", Optics Express, **14**, 11242-11247 2006).
3. D. Ceus, A. Tonello, L. Grossard, L. Delage, F. Reynaud, H. Herrmann, W. Sohler, "Phase closure retrieval in an infrared-to-visible upconversion interferometer for high resolution astronomical imaging", Optics express, **19**, 8616-8624 (2011).
4. O. Tadanaga, T. Yanagawa, Y. Nishida, H. Miyazawa, "Efficient 3-µm difference frequency generation using direct-bonded quasi-phase-matched LiNbO3 ridge waveguides", Appl. Phys. Lett., **88**, 061101 (2006).
5. A. Jechow, M. Schedel, S. Stry, J. Sacher, R. Menzel, "Highly efficient single-pass frequency doubling of a continuous-wave distributed feedback laser diode using a PPLN waveguide crystal at 488 nm", Optics Letters, **32**, 3035-3037 (2007).
6. J. Leuthold, C. Koos, W. Freude, Nonlinear silicon photonics, Nature Photonics **4**, 535 - 544 (2010)
7. D. J. Moss, R. Morandotti, A. L. Gaeta, and M. Lipson,"New CMOS-compatible platforms based on silicon nitride and Hydex for nonlinear optics," Nature Photonics **7**, 597-607 (2013)
8. M. L. Bortz, M. A. Arbore, and M. M. Fejer, "Quasi-phase-matched optical parametric amplification and oscillation in periodically poled LiNbO3 waveguides", Opt. Lett. , **20**, 49-51 (1995).
9. F Pettazzi, V Coda, M Chauvet, E Fazio, "Frequency-doubling in self-induced waveguides in lithium niobate", Optics communications **272** , 238-241
10. K. R. Parameswaran, R. K. Route, J. R. Kurz, R. V. Roussev, M. M. Fejer, and M. Fujimura, "Highly efficient second-harmonic generation in buried waveguides formed by annealed and reverse proton exchange in periodically poled lithium niobate", Optics Letters, **27**, pp. 179-181 (2002)
11. R. Kou, S. Kurimura, K. Kikuchi, A. Terasaki, H. Nakajima, K. Kondou, J. Ichikawa, "High-gain, wide-dynamic-range parametric interaction in Mg-doped LiNbO3 quasi-phase-matched adhered ridge waveguide", Optics Express, **19**, 11867 (2011)
12. S. Kurimura, Y. Kato, M. Maruyama, Y. Usui, H. Nakajima, "Quasi-phase-matched adhered ridge waveguide in LiNbO3", Applied Physics Letters **89**, 191123 (2006);
13. Y. Shibata, K. Kaya, K. Akashi, M. Kanai, T. Kawai and S. Kawai, "Epitaxial growth of LiNbO3 thin films by excimer laser ablation method and their surface acoustic wave properties", Appl. Phys. Lett. **61**, 1000 ( 1992)
14. J. H. Haeni, P. Irvin, W. Chang, R. Uecker, P. Reiche, Y. L. Li, S. Choudhury, W. Tian, M. E. Hawley, B. Craigo, A. K. Tagantsev, X. Q. Pan, S. K. Streiffer, L. Q. Chen, S. W. Kirchoefer, J. Levy, and D. G. Schlom, Nature (London) **430**, 758 (2004).
15. P. Rabiei, P. Günter, "Optical and electro-optical properties of submicrometer lithium niobate slab waveguides prepared by crystal ion slicing and wafer bonding", Applied Physics Letters, **85**, 4603 (2004)

16. Organic thin films for waveguiding nonlinear optics, Edited by F. Kajzar and J.D. Swalen, advances in nonlinear optics, CRC Press, P181 (1996).
17. K. C. Rustagi, S.C. Mehendale, S. Meenakshi, "optical frequency conversion in quasi-phase-matched stacks of nonlinear crystals", IEEE journal of quantum electronics, **18**, 1029-1041, (1982)
18. L. E. Myers, R. C. Eckardt, M. M. Fejer, R. L. Byer, W. R. Bosenberg, and J. W. Pierce, "Quasi-phase-matched optical parametric oscillators in bulk periodically poled LiNbO3", JOSAB, **12**, 2102-2116 (1995).
19. E. Courjon, N. Courjal, W. Daniau, G. Lengaigne, L. Gauthier-Manuel, S. Ballandras, and J. Hauden, "Lamb wave transducers built on periodically poled Z-cut LiNbO3 wafers," J. Appl. Phys. **102**, 114107 (2007).
20. J. H. Park, T.Y. Kang, J.H. Ha, and H.Y. Lee, "Spatial mode behavior of second harmonic generation in a ridge-type waveguide with a periodically poled MgO-doped lithium niobate crystal", Japanese Journal of Applied Physics, **53**, 062201 (2014)
21. F. Devaux, E. Lantz, M. Chauvet, "3D-PSTD for modelling second harmonic generation in periodically poled lithium niobate ridge-type waveguides" JOSAB (accepted for publication).
22. F. Pignatiello, M. De Rosa, P. Ferraro, S. Grilli, P. De Natale, A. Arie, S. De Nicola, "Measurement of the thermal expansion coefficients of ferroelectric crystals by a moiré interferometer", Optics Communications, **277**, 14-18 (2007).

## 1. Introduction

Optical wavelength conversion is a key function for various domains of application. For instance, in telecommunication networks it is useful for wavelength division multiplexing systems [1,2], in astronomy it can be used to transfer optical signals to wavelengths where efficient detectors are available [3] or more generally it can offer optical sources at particular wavelengths [4,5]. In the push for silicon photonics technology, nonlinear optical functionalities also have to be conceived. Remarkable results have been obtained in silicon nanowaveguides [6] or in other highly nonlinear materials [7] that directly benefit from the mature technological equipment developed for microelectronics. However, some drawbacks are inherent to the use of the third order nonlinearities exploited in these latter materials such as, the need for high optical power and the detrimental presence of multiphoton absorption at telecommunication wavelengths. An alternative solution for all-optical information processing based on silicon is to realize hybrid structures using high nonlinear properties of well-known materials such as $LiNbO_3$ components.

$LiNbO_3$ which is widely used in the photonics industry has among one of the highest quadratic nonlinear coefficient. It thus make it a material of choice to develop frequency conversion systems and more generally nonlinear optical functions [8,9]. Moreover, $LiNbO_3$ can be periodically poled to allow quasi-phase matching for various combination of wavelengths in its transparency window (0.4μm-5μm). High conversion efficiency with low power beams can even be reached in waveguides thanks to light confinement. While conventional fabrication techniques such as titanium in-diffusion or proton exchange can form gradient index waveguides with great performances [3,5,10], recent developments have led to the fabrication of ridge waveguides that provides more flexibility and more stable behavior even under high average optical power [4,11,12]. For this latter configuration, initial fabrication stage consists in the realization of thin periodically poled $LiNbO_3$ (PPLN) films on a chosen substrate. Epitaxial growth of ferroelectrics films is a field in development but it does not yet provide the required crystal quality for applications [13-14]. However, when submicron thicknesses are required, films can be formed from bulk $LiNbO_3$ by the smart-cut technique [15] also called ion slicing. Another technique adequate for thicker layers is based on grinding and polishing. Ultimately 2-D waveguides are formed by etching or precision dicing of these films [4, 11-12]. With this method, hybrid structures with high index contrast can give waveguides with extreme confinement on a variety of substrates. When films are realized by micromechanical processes, properties of bulk $LiNbO_3$ material are retained, notably its high nonlinear coefficient. In addition, compare to standard fabrication based on photolithography, clean room facilities for the fabrication. In the present work, we describe the realization of ridge PPLN waveguides on silicon wafers. Moreover, the components are entirely carved with a precision sawing machine without need for grinding or polishing steps which constitutes an original feature. The components are designed for efficient second harmonic generation (SHG) at telecommunication wavelengths.

## 2. Modeling

To determine the poling period to achieve quasi-phase matching for second harmonic in a ridge PPLN waveguide at telecom wavelengths $\lambda_p$ a commercial software is used. For a given waveguide structure the effective indices $n^{eff}$ of the fundamental modes are determined. Quasi-phase matching is obtained when the poling period $\Lambda$ satisfies the following relation between effective indices of the pump, $n_\omega$ and of the SHG signal, $n_{2\omega}$ :

$$\Lambda = \frac{\lambda_p}{2(n_{2\omega} - n_\omega)} \quad (1)$$

Figure 1 presents the calculated poling period versus the dimension of a square ridge congruent LiNbO$_3$ waveguide separated from a silicon substrate by silica layer. Three curves are presented for pump at the telcom wavelengths of 1500nm, 1550nm and 1600nm. In order to benefit from the high $d_{33}$ non-linear coefficient associated with an extraordinary polarized light in LiNbO$_3$ TM polarizations are considered for both the pump and the SHG signal. Moreover, even though waveguides can be multimode, only phase matching between fundamental guided modes are taken into account.

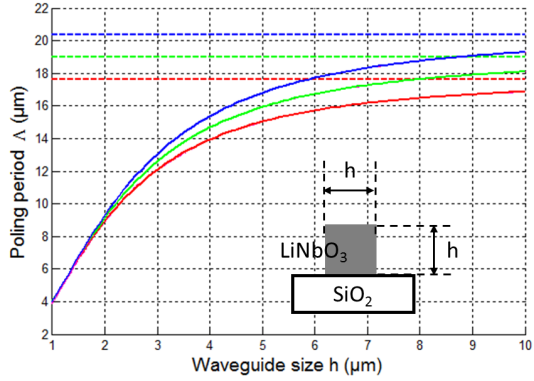

Fig. 1. Poling period for quasi-phase-matched SHG between fundamental modes in a congruent LiNbO$_3$ square ridge waveguide on silica for three different pump wavelengths at room temperature (continuous line from bottom to top : $\lambda_p$=1500nm, 1550nm and 1600nm). Dotted horizontal lines indicate poling period at the three wavelengths in bulk LiNbO$_3$.

We observe in figure 1 that the poling period for quasi-phase-matching becomes smaller as the waveguide section is decreased. For the present study fabricated waveguides have cross sections of about 8µm by 8µm which correspond to a poling period $\Lambda$ of 17.5µm for a pump wavelength at 1550nm. Waveguides with smaller cross section are expected to provide higher SHG conversion efficiencies thanks to a better confinement but it is at the expense of more stringent conditions for fabrication and for light coupling. This choice of waveguide geometry is thus a trade-off in order to release some tolerances for the fabrication and also facilitate light coupling with optical fibers.

To characterize the second harmonic effect it is usual to determine the normalized conversion efficiency $\eta_{SHG}$ (%/W) defined in the low conversion regime (undepleted pump) in a waveguide of length $l$ for a quasi-phase-matched configuration [16,17].

$$\eta_{SHG} = \frac{p_{2\omega}}{p_\omega^2} = \frac{4}{\pi c^3 \varepsilon_0} \frac{\Gamma \omega^2 d_{eff}^2 l^2}{s\, n_{2\omega} n_\omega^2} Sinc^2\left(\frac{\Delta k\, l}{2}\right) \quad (2)$$

Where $P_\omega$ and $P_{2\omega}$ are the pump and SHG signal power, respectively. Γ is the overlap integral of the SHG and pump mode, $d_{eff}$ is the effective nonlinear coefficient equal to $d_{33}$, $S$ is the effective cross section of the waveguide and the phase mismatched is given by :

$$\Delta k = \frac{4\pi (n_{2\omega} - n_\omega)}{\lambda_p} \qquad (3)$$

To reach high $\eta_{SHG}$ values the main challenge is to maintain the phase-matching condition ($\Delta k = 0$) over the length of the waveguide. For a given poling period, the modes effective indices have to be invariant along propagation. As a consequence, a waveguide with a uniform geometry has to be fabricated. While the poling period is precisely set by the mask period used during the poling process, the waveguide geometry can to the contrary suffer from fabrication non uniformities. To have some insight on the impact of waveguide nonuniformity on the SHG process, the coupled wave equation that gives the evolution of $a_{2\omega}$ and $a_\omega$ the field amplitude of the SHG and pump beams, respectively, along propagation distance $z$ is solved by the Runge-Kutta method.

$$\frac{da_\omega}{dz} = -jg a_\omega^* a_{2\omega} \exp(-j\Delta k\, z)$$
$$\frac{da_{2\omega}}{dz} = -jg\, a_\omega a_{2\omega} \exp(-j\Delta k\, z) \qquad (4)$$

Where the coupling coefficient is $\quad g = \dfrac{2\omega\, d_{eff}}{c\, n_{2\omega}} \qquad (5)$

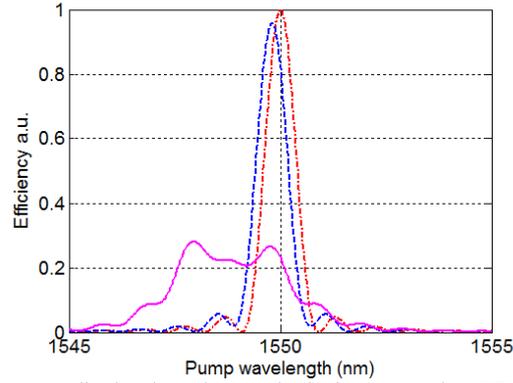

Fig. 2. Conversion efficiency normalized to the optimum value in three 15 mm long PPLN ridges with a poling period set for conversion at 1550nn and central cross section of 7.5µm*7.5µm. Dotted-dashed lines (ideal structure); dashed curve : 50nm thickness error; continuous line : 300nm thickness error.

The commercial software COMSOL is first used to calculate the modes effective indices as a function of the waveguide cross section and wavelengths in order to calculate $\Delta k$. $\Delta k$ can vary with arbitrary law along propagation since it depends on the effective indices of the guided modes according to eq. (3). We assume slow variations of the cross sections which authorize to write $n_{eff}$ as being linearly dependent of waveguide dimensions and wavelengths. Then the system of equation (4) is solved to deduce the conversion efficiency. Figure 2 presents the predicted conversion efficiency in a 15mm long waveguides with a varying thicknesses apart from a central 7.5µm square side section. The poling period is set to give a maximum conversion at 1550nm for the perfect waveguide. For the latter case, an ideal sinus cardinal response is found (dotted-dashed curve) as expected from eq. (2). If the waveguide thickness varies linearly from 7.475µm at the input to 7.525µm at the output, that is a 50nm change, the maximum conversion and spectral response are slightly altered (dashed curve). If the thickness linear variation is increased to 300nm the spectral response is enlarged and distorted with two peaks and a conversion efficiency 4 times weaker than in a perfect waveguide. To show the overall behavior the evolution of the maximum conversion efficiency and the width of the spectral response has been computed in Figure 3 for a waveguide thickness non uniformity up to 1μm. To obtain a conversion efficiency better than half the optimum conversion for the studied structure, we deduce that the thickness variation of the waveguide should be less than 150nm.

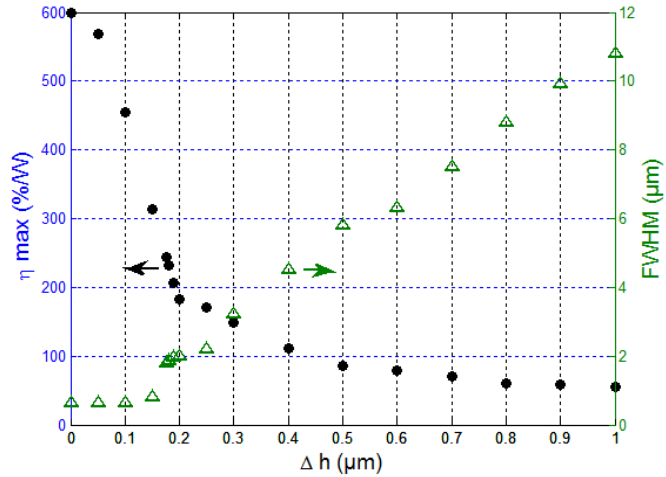

Fig. 3. Calculated maximum (dots) and FWHM (triangles) of the SHG conversion efficiency response versus fabrication error of the waveguide thickness. A 15mm long PPLN ridge waveguide whose central cross section is 7.5μm*7.5μm with a poling period set for conversion at 1550nm is considered.

## 3. Fabrication

The first stage of the fabrication is to periodically pole a 500μm thick commercial z-cut congruent $LiNbO_3$ wafer, supplied by Gooch & Housego, by application of an intense electric field at room temperature [18, 19]. In a second stage, a typical 300nm thick $SiO_2$ layer is deposited by ICPECVD onto one face of the poled wafer followed by the sputtering of a 200nm thick gold layer. A high flatness silicon wafer is also coated with a 200nm thick gold layer. The metalized faces of both the PPLN and silicon wafers are then placed into contact and pressed in a wafer bonding machine. The bonding process is realized at room temperature which prevents mechanical stress that could occur due to the dissimilar temperature coefficients of the two wafers. At this stage, a typical 1 mm thick hybrid structure composed of a silicon substrate bonded to a PPLN wafer is obtained. A possible solution to form a thin $LiNbO_3$ layer is to thin down the heterostructure by grinding and polishing. This method was used earlier to realize adhered nonlinear ridge waveguides [11, 12] where epoxy was used to bond PPLN wafers onto $LiNbO_3$ substrates. This process leads to components with remarkable characteristics. Multiple nonlinear components with a common thickness can thus be realized on the same substrate. In order to extend this work to silicon photonics compatible components and at the same time gain some flexibility we developed a new process to fabricate ridge waveguides.

A precision dicing saw (model DISCO DAD3350) is used to entirely carve the ridge waveguides from the hybrid wafer. 56 mm diameter diamond blades, 400μm thick, are employed. To realize a waveguide the first step consists in thinning locally the PPLN material down to the desired thickness. It thus requires to precisely locate the level of the blade tip which is accomplished with an accuracy better than 100nm thanks to an initial calibration procedure. The second step consists in dicing two parallel trenches inside the thinned area in order to form the lateral sides of the ridge waveguide. At last, the hybrid wafer is diced to give polished input and output faces. Figure 4 presents a SEM image of a 10µm by 10µm cross section waveguide realized with this technique. Both sides of the ridge are curved due the small radius of curvature of each corner of the blade. By optical profilometry a 3-4 nm RMS roughness is measured for the ridges faces. Such a surface quality ensures low propagation losses. Moreover, no further input or output faces polishing is necessary. One additional advantage of using the dicing technique to realize the ridge compared to the polishing technique is to avoid the appearance of some periodic undulation at the surface of the ridge. Indeed,–z and +z faces have disparate mechanical properties. To be more specific, -z faces are harder than +z face. As a consequence, when PPLN surface is polished a periodic corrugation given by the periodic poling appears at the surface. This notably increases the propagation losses and can even give parasitic Bragg diffraction. When the top surface of the ridge is instead cut with the precision sawing machine no such corrugation is present. This is illustrated in Figure 5 where line profiles have been taken with an optical profilometer for ridge waveguides realized either by polishing or by precision dicing. An undulation of 40nm amplitude with a period of about 20µm clearly appears at the surface of the polished waveguide while precision dicing gives a flatter surface.

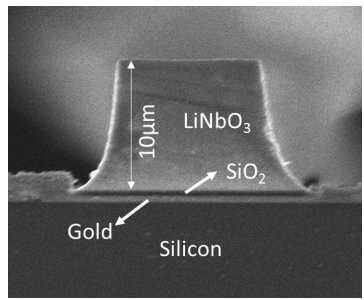

Fig. 4. SEM image of a 10µm by 10µm ridge waveguide realized by precision dicing.

15 ridge waveguides where diced from the above described hybrid wafer. All waveguides are 15 mm long but different cross sections are targeted to demonstrate the versatility of the developed fabrication technique but also to see their influence on the SHG process. Figure 6 presents a selection of realized ridge waveguides. Fig 6a shows a 7µm square section waveguide. Similar section but with more vertical side walls can be obtained by dicing deeper in the silicon substrate (fig. 6b). Different aspect ratio waveguides can also be realized as witnessed by the 12µm high by 5µm large ridge in Figure 6c or the 2.6µm thick by 10µm large ridge in Figure 6d. Note that all realized waveguides are functional which demonstrate the excellent adherence of the bonding and the great potential of the fabrication technique.

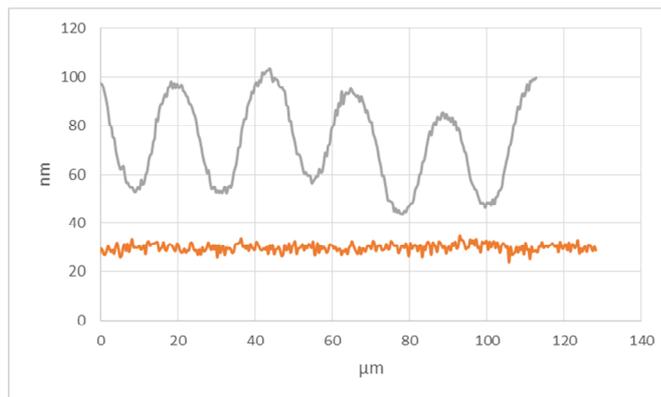

Fig. 5. Profile measured at the top surface of a PPLN ridge with an optical profilometer for a waveguide thinned by polishing (curve with undulations) or by precision dicing (flat curve).

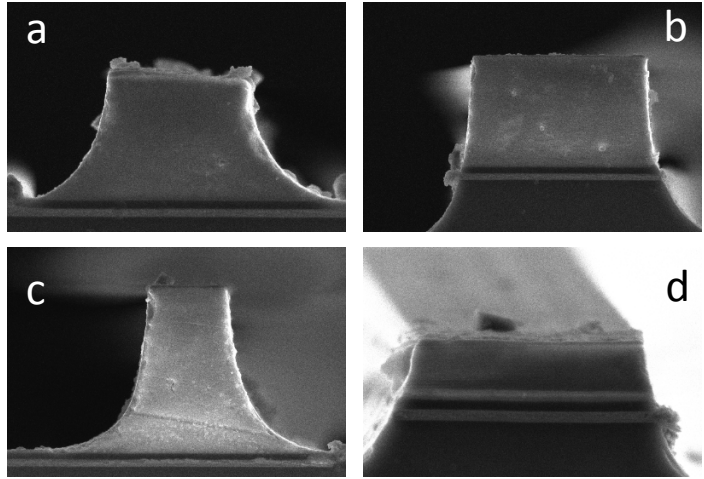

Fig. 6. SEM images of four ridge waveguides diced from the same hybrid wafer.

## 4. Optical characterization

An optical set-up is then assembled to characterize the waveguides for SHG generation. The optical source is an external cavity CW laser tunable between 1500nm and 1600nm that delivers the pump beam. The maximum output power is 10mW. The laser output is attached to a lensed fiber to form a spatially collimated beam. The light is linearly polarized and focused with an antireflection coated microscope objective at the input face of the waveguide under test. The waveguides is placed on a temperature regulated mount controlled by a Peltier element. Output light is then imaged via a second microscope objective onto a vidicon camera to optimize the coupling. A longpass dichroic mirror is placed at the output in order to separate the SHG signal from the unconverted pump beam. Each beam is measured separately either with powermeters or onto cameras to analyze the guided spatial mode distribution. This set-up allows measurement of the conversion efficiency of the SHG process and is also used to evaluate the transmittivity of the waveguide. Propagation losses at 1550nm are evaluated to be 0.8dB for the TM mode.

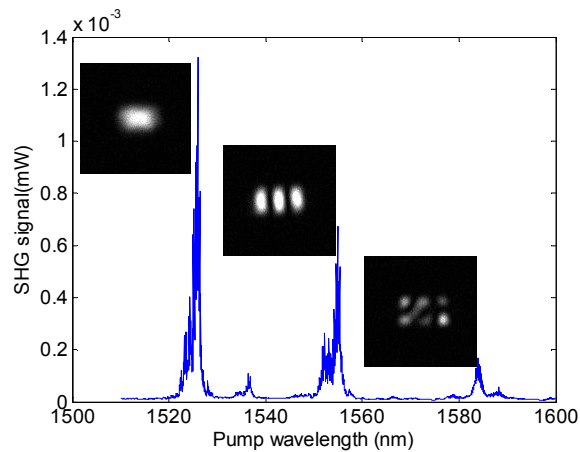

Fig. 7. Measured second harmonic signal in a ridge waveguide versus pump wavelength. Insets show the observed mode distribution corresponding to the three main peaks.

SHG spectral response of the PPLN waveguides are first analyzed over the whole spectrum of the laser source. As an example, the amplitude of the detected SHG signal versus pump wavelength is presented in figure 7 for the 12µm by 8.5µm waveguide depicted in figure 6b with a poling period of 17.5µm. Several second harmonic peaks are clearly seen in the analyzed spectral range. To better comprehend this spectral response, light distribution at the output of the waveguide is monitored with a CCD camera that is only sensitive to SHG signal. Each peak is linked to a different transverse mode profile as can be seen in the insets in figure 7. The strongest peak at 1526nnm corresponds to the fundamental $TM_{00}$ mode while other peaks are higher order modes of the waveguide. Indeed since the waveguide is multimode phase matching conditions can be fulfilled for different mode combinations as shown in ref [20]. The highest conversion efficiency is reached when two photons from the $TM_{00}$ pump mode combine to give a photon in the $TM_{00}$ SH mode thanks to an optimized mode overlap. Excitation of higher order modes can be minimized by carefully adjusting the size and the alignment of the injected beam [21]. Consequently, the energy of the guided beam mainly propagates in the fundamental mode which is favorable to obtain high conversion for frequency doubling. Note that, for the considered waveguide, the measured phase matching wavelength of 1526nm is very close the one predicted in figure 1 for a square waveguide with similar cross section (10*10µm$^2$ ) and with the same poling period of 17.5µm.

Spectral response of SHG corresponding to $TM_{00}$ modes is then considered further in the text. Normalized conversion efficiency obtained in the waveguide corresponding to Fig. 6a (Λ = 17.5µm) is depicted in figure 8 for a temperature of 24°C. This measurement was performed with waveguide input and output faces antireflection coated for pump wavelength and SHG signal, respectively, in order to limit perturbations due to the Fabry-Pérot effect. The optimum wavelength is 1545.9nm and a 1nm FWHM is measured for the central peak width which has to be compared of the one estimated (0.65nm) for a perfect waveguide of similar cross section (figure 3). A linear change of the waveguide thickness of 150nm over the component length could explain this wider response. Moreover, the experimental response is observed to be asymmetric relative to the central wavelength. This behavior cannot be explained if a simple linear variation of the waveguide dimension is considered. Additional simulations using eq. (4) reveal that a waveguide dimension varying with a arbitrary parabolic law of amplitude 50nm in addition to the 150nm linear variation gives a response close to the observed one. We can conclude that the fabrication technique is able to produce waveguides with good uniformity.

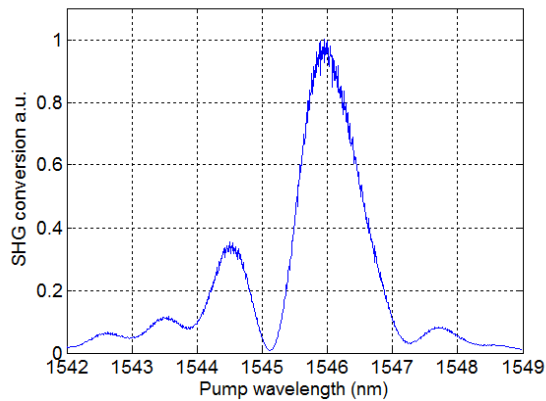

Fig. 8. SHG Conversion efficiency normalized to unity measured in waveguide from Fig.6a at 24°C.

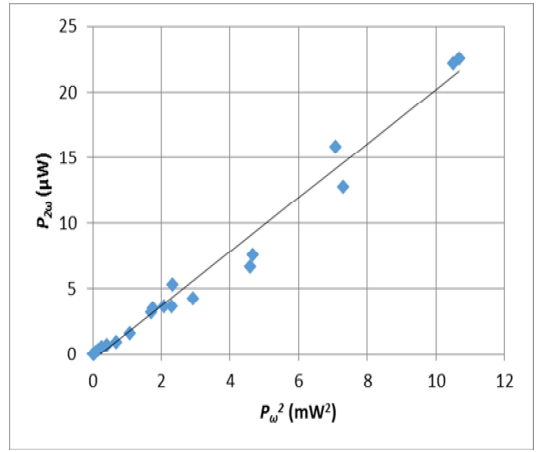

Fig. 9. Power of SHG signal versus squared power of pump beam measured in the waveguide corresponding to Figure 6a.

To corroborate this assertion, the maximum conversion efficiency $\eta_{SHG}$ of the component was evaluated in the low conversion regime by measuring the output SH beam power versus the pump power. The waveguide is kept at 24°C and the pump wavelength is set at the optimum value of 1545.9nm. A linear evolution of the SHG signal versus the square of the injected pump is measured as depicted in figure 9. The slope of the curve gives a conversion efficiency $\eta_{SHG}$ of 204%/W. This maximum conversion efficiency is consistent with a variation of the waveguide thickness of about 150-200 nm according to figure 3. Additional simulations, not shown here, indicate that the non uniformities of the waveguide width or thickness have very similar influence on the conversion efficiency. As a first approach, a constant cross section has to be maintained for an optimized conversion. A straightforward way to improve the component efficiency is thus to better control the fabrication process to reach a more uniform geometry. Studies are in progress to correct or compensate the small variations of the geometry by a post-fabrication processs while the conversion efficiency is analyzed in real time.

While previous measurements have been performed with a component kept at room temperature, we may wonder if the component can sustain temperature changes. This characteristic would bring fine tuning of the convertion wavelength. Such a question arises since there is a large difference between the thermal expansion coefficients of silicon (2.6 $10^{-6}$ °C$^{-1}$) and LiNbO$_3$ (13 $10^{-6}$ °C$^{-1}$) [22]. The temperature controller based on a Peltier element allows sample temperature stabilization in the 15°C to 60°C range. Figure 10 presents the measured optimum converted wavelength versus temperature for the previously tested waveguide from fig. 6a and for a bulk PPLN sample with a poling period of 18.6μm. No component failure occurs even though several temperature cycles were performed between room temperature and 60°C. The thin and narrow waveguides carved on the same silicon substrate thus do not suffer excessive strain over this temperature range. Note that a 500μm thick silicon substrate gold bonded to a 500μm thick LiNbO$_3$ wafer breaks under the same temperature test. The converted wavelengths shown in Figure 10 are in accordance with the prediction from figure 1. A longer PPLN period of 18.6μm is shown for the bulk PPLN since a similar period than the waveguide under test (17.5μm) would not give any conversion for the 1500-1600nm range. The shift in wavelength is clearly observed. Furthermore, a linear fit of the measurements presented in figure 10 gives a wavelength variation of 0.126 nm/°C for the waveguide in comparison with a value of 0.138 nm/°C in the bulk PPLN medium. This slightly lower temperature dependence can be attributed to the influence of the dispersive effect of the waveguide.

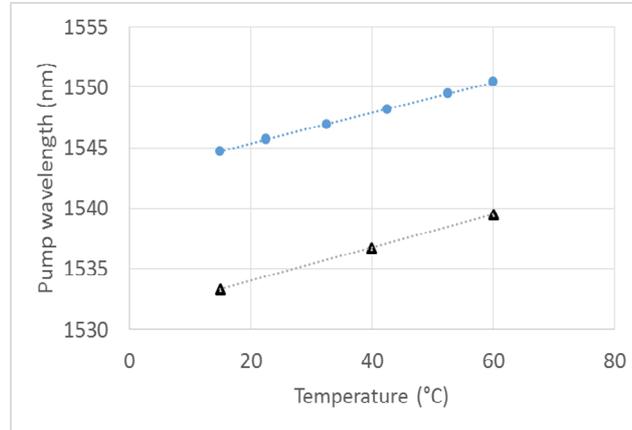

Fig. 8. Optimum pump wavelength versus sample temperature. Dots : PPLN waveguide from figure 6a (poling period 17.5μm) . Triangles : bulk PPLN with a poling period of 18.6μm.

## 5. Conclusions

To conclude, we have shown that PPLN ridge waveguides for nonlinear optical conversion can be fabricated onto silicon substrate by wafer bonding followed by direct precision dicing. Dicing produces waveguides with low roughness and avoids periodic surface corrugation due to poling that appears when polishing is instead used to thin down the wafer. Influence of geometrical dimension of waveguides is found to be the critical parameter in order to maintain a good phase matching for nonlinear optical process. Best waveguides give high second harmonic conversion efficiency up to 204%/W measured at telecommunication wavelengths in 16 mm long devices. Study of temperature dependence of components demonstrates that PPLN waveguides are robust and offer tunability of converted wavelengths similar than in bulk PPLN.

Financial support by the European Program ActPhast and by the Labex ACTION program (contract No. ANR-11-LABX-0001-01) are gratefully acknowledged. This work was partly supported by the RENATECH network and its FEMTO-ST MIMENTO technological facility.